\documentclass{article}
\usepackage{lscape}
\usepackage{epsfig}

\begin{document}
\title{Electric dipoles at ultralow temperatures}
\author{John L. Bohn}
\maketitle

\section{General remarks}

Any object with a net positive charge on one end and a net negative
charge on the other end possesses an electric  dipole moment.  In
ordinary classical electromagnetism this dipole moment is a vector
quantity that can point in any direction, and is subject to
electrical forces that are fairly straightforward to formulate
mathematically. However, for a quantum mechanical object like an
atom or molecule, the strength and orientation of the object's
dipole moment can depend strongly on the object's quantum mechanical
state.  This is a subject that becomes relevant in low temperature
molecular samples, where an ensemble of molecules can be prepared in
a single internal state, as described elsewhere in this volume. In
such a case, the mathematical description becomes more elaborate,
and indeed the dipole-dipole interaction need not take the classical
form given in textbooks. The description of this interaction is the
subject of this chapter.

We approach this task in three steps: first, we introduce the {\it
ideas} of how dipoles arise in quantum mechanical objects; second,
we present a {\it formalism} within which to describe these dipoles;
and third, we give {\it examples} of the formalism that illustrate
some of the basic physics that emerges. The discussion will explore
the possible energy states of the dipole, the field generated by the
dipole, and the interaction of the dipole with another dipole.  We
restrict the discussion to a particular ``minimal realistic model,''
so that the most important physics is incorporated, but the
arithmetic is not overwhelming.

Although we discuss molecular dipoles in several contexts, our main
focus is on polar molecules that possess a $\Lambda$ doublet in
their ground state.  These molecules are the most likely, among
diatomic molecules at least, to exhibit their dipolar character at
moderate laboratory field strengths.  $\Lambda$-doubled molecules
have another peculiar feature, namely, their ground states possess a
degeneracy even in an electric field.  This means that there is more
than one way for such a molecule to align with the field; the two
possibilities are characterized by different angular momentum
quantum numbers.  This degeneracy leads to novel properties of both
the orientation of a single molecule's dipole moment and the
interaction between dipoles. In the examples we present, we focus on
revealing these novel features.

We assume the reader has a good background in undergraduate quantum
mechanics and electrostatics. In particular, the ideas of matrix
mechanics, Dirac notation, and time-independent perturbation theory
are used frequently. In addition, the reader should have a passing
familiarity with electric dipoles and their interactions with fields
and with each other. Finally, we will draw heavily on the
mathematical theory of angular momentum as applied to quantum
mechanics, as described in Appendix A of this volume, and in more
detail in the classic treatise of Brink and Satchler ~\cite{BS}.
When necessary, details of the structure of diatomic molecules have
been drawn from Brown and Carrington's recent authoritative text
\cite{BC}.

\section{Review of Classical Dipoles}

The behavior of a polar molecule is largely determined by its
response to electric fields.  Classically, an electric dipole
appears when a molecule has a little bit of positive charge
displaced a distance from a little bit of negative charge.  The
dipole moment is then a vector quantity that characterizes the
direction and magnitude of this displacement:
\begin{eqnarray}
{\vec \mu} = \sum_{\xi} q_{\xi} {\vec r}_{\xi} , \nonumber
\end{eqnarray}
where the $\xi^{\rm th}$ charge $q_{\xi}$ is displaced ${\vec
r}_{\xi}$ from a particular origin. Because we are interested in
forces exerted on molecules, we will take this origin to be the
center of mass of the molecule. (Defining ${\vec \mu}=0$ would
instead identify the center of charge of the molecule --  quite a
different thing!)  By convention, the dipole moment vector points
away from the negative charges, and toward the positive charges,
inside the molecule.

A molecule has many charges in it, and they are distributed in a
complex way, as governed by the quantum mechanical state of the
molecule. In general there is much more information about the
electrostatic properties of the molecule than is contained in its
dipole moment. However, at distances far from the molecule (as
compared to the molecule's size), these details do not matter.  The
forces that one molecule exerts on another in this limit is strongly
dominated by the dipole moments of the two molecules. We consider in
this chapter only electrically neutral molecules, so that the
Coulomb force between molecules is absent. In this limit, too, the
details of the dipole moment's origin are irrelevant, and we
consider the molecule to be a ``point dipole,'' whose dipole moment
is characterized by a magnitude $\mu$ and a direction ${\hat \mu}$.

If a dipole ${\vec \mu}$ is immersed in an electric field ${\vec
{\cal E}}$, its energy depends on the relative orientation of the
field and the dipole, via
\begin{eqnarray}\label{dipole_field_Hamiltonian}
E_{el} = -{\vec \mu} \cdot {\vec {\cal E}}. \nonumber
\end{eqnarray}
 This follows simply from the fact that the
positive charges will be pulled in the direction of the field, while
the negative charges are pulled the other way. Thus a dipole
pointing in the same direction as the field is lower in energy than
a dipole pointing in exactly the opposite direction. In classical
electrostatics, the energy can continuously vary between these two
extreme limits.

As an object containing charge, a dipole generates an electric
field, which is given, as usual, by the gradient of an electrostatic
potential, ${\vec {\cal E}}_{\rm molecule} = - {\vec \nabla} \Phi
({\vec r})$. For a point dipole the potential $\Phi$ is given by
\begin{eqnarray} \label{classical_dipole}
\Phi({\vec r}) = { {\vec \mu} \cdot {\hat r} \over r^2},
\end{eqnarray}
where ${\vec r} = r{\hat r}$ denotes the point in space, relative to
the dipole, at which the field is to be evaluated \cite{Jackson}.
The dot product in (\ref{classical_dipole}) gives the field $\Phi$ a
strong angular dependence.  For this reason, it is convenient to use
spherical coordinates to describe the physics of dipoles, since they
explicitly record directions.  Setting ${\hat r} = (\theta, \phi)$
and ${\hat \mu} = (\alpha, \beta)$ in spherical coordinates, the
dipole potential becomes
\begin{eqnarray} \label{dot_product}
\Phi = {\mu \over r^2} \left( \cos \alpha \cos \theta + \sin \alpha
\sin \theta \cos (\beta - \phi) \right).
\end{eqnarray}
For the most familiar case of a dipole aligned along the positive
$z$-axis ($\alpha = 0$), this yields the familiar result $\Phi = \mu
\cos \theta / r^2$.  This potential is maximal along the dipole's
axis ($\theta=0$ or $\pi$), and vanishes in the direction
perpendicular to the dipole ($\theta = \pi/2$).

From the results above we can evaluate the interaction potential
between two dipoles.  One of the dipoles generates an electric
field, which acts on the other.  Taking the scalar product of one
dipole moment with the gradient of the dipole potential
(\ref{dot_product}) due to the other, we obtain \cite{Jackson}
\begin{eqnarray}
\label{dd_interaction} V_d ({\vec R}) = { {\vec \mu}_1 \cdot {\vec
\mu}_2 - 3 ({\vec \mu}_1 \cdot {\hat R})  ({\vec \mu}_2 \cdot {\hat
R} ) \over R^3 }, \nonumber
\end{eqnarray}
where ${\vec R} = R{\hat R}$ is the relative coordinate of the
dipoles. This result is general for any orientation of each dipole,
and for any relative position of the pair of dipoles. In a special
case where both dipoles ar aligned along the positive $z$-axis, and
where the vector connecting the centers-of-mass of the two dipoles
makes an angle $\theta$ with this axis, the dipole-dipole
interaction takes a simpler form:
\begin{eqnarray}
\label{pure_dipole_interaction} V_d ({\vec r}) = \mu_1 \mu_2 {1 - 3
\cos^2 \theta \over R^3}.
\end{eqnarray}
Note that the angle $\theta$ as used here has a different meaning
from the one in Eq. (\ref{dot_product}).  We will use $\theta$ is
both contexts throughout this chapter, hopefully without causing
undue confusion.   The form (\ref{pure_dipole_interaction}) of the
interaction is useful for illustrating the most basic fact of the
dipole-dipole interaction: if the two dipoles line up in a
head-to-tail orientation ($\theta = 0$ or $\pi$), then $V_d<0$ and
they attract one another; whereas if they lie side-by-side
($\theta=\pi/2$), then $V_d>0$ and they repel one
another.\footnote{This expression ignores a contact potential that
must be associated to a point dipole to conserve lines of electric
flux \cite{Jackson}. However, real molecules are {\it not} point
dipoles, and the electrostatic potential differs greatly from this
dipolar form at length scales inside the molecule, scales that do
not concern us here.}

Our main goal in this chapter is to investigate how these classical
results change when the dipoles belong to molecules that are
governed by quantum mechanics.  In Sec. 3 we evaluate the energy of
a dipole exposed to an external field; in Sec. 4 we consider the
field produced by a quantum mechanical dipole; and in Sec. 5 we
address the interaction between two dipolar molecules.

\section{Quantum mechanical dipoles in fields}

Whereas the classical energy of a dipole in a field can take a
continuum of values between its minimum and maximum, this is no
longer the case for a quantum mechanical molecule.  In this section
we will establish the spectrum of a polar molecule in an electric
field, building from a set of simple examples.  To start, we will
define the laboratory $z$-axis to coincide with the direction of an
externally applied electric field, so that ${\vec {\cal E}} = {\cal
E}{\hat z}$. In this case, the projection of the total angular
momentum on the $z$-axis is a conserved quantity.

\subsection{Atoms}

Our main focus in this chapter will be on electrically polarizable
dipolar molecules.  But before discussing this in detail, we first
consider the simpler case of an electrically polarizable atom,
namely, hydrogen.  This will introduce both the basic physics ideas,
and the angular momentum techniques that we will use.  In this case
a negatively charged electron separated a distance $r$ from a
positively charged proton forms a dipole moment ${\vec \mu} =
-e{\vec r}$.

Because dipoles require us to consider directions, it is useful to
cast the unit vector ${\hat r}$ into its spherical components
\cite{BS}:
\begin{eqnarray}
\left({x  \over r} \pm i{y \over r}\right) = \mp \sqrt{2}  C_{1 \pm
1} (\theta, \phi), \;\;\;\; {z \over r} =  C_{10}(\theta, \phi).
\nonumber
\end{eqnarray}
Here the $C$'s are reduced spherical harmonics, related to the usual
spherical harmonics by \cite{BS}
\begin{eqnarray}
C_{kq} = \sqrt{ { 4 \pi \over 2k+1} } Y_{kq}, \nonumber
\end{eqnarray}
and given explicitly for $k=1$ by (Appendix A)
\begin{eqnarray}
\label{C_functions} C_{1\pm 1}(\theta\phi) = \mp {1 \over \sqrt{2}}
\sin \theta e^{\pm i \phi} \;\;\;\; C_{10}(\theta \phi) = \cos
\theta.
\end{eqnarray}
In general, it is convenient to represent interaction potentials in
terms of the functions $C_{kq}$ (since they do not carry extra
factors of $4\pi$), and to use the functions $Y_{kq}$ as wave
functions in angular degrees of freedom (since they are already
properly normalized, by $\langle Y_{lm} | Y_{l^{\prime}m^{\prime}}
\rangle = \delta_{ll^{\prime}} \delta_{mm^{\prime}}$).  Integrals
involving the reduced spherical harmonics are conveniently related
to the 3-$j$ symbols of angular momentum theory, for example:
\begin{eqnarray}
&& \int d(\cos \theta) d \phi C_{k_1q_1} (\theta \phi)  C_{k_2q_2}
(\theta \phi) C_{k_3q_3} (\theta \phi) \nonumber \\ &&= 4\pi \left(
\begin{array}{ccc}
k_1 & k_2 & k_3 \\
q_1 & q_2 & k_3
\end{array} \right)
\left(
\begin{array}{ccc}
k_1 & k_2 & k_3 \\
0 & 0 & 0
\end{array} \right). \nonumber
\end{eqnarray}
The 3-$j$ symbols, in parentheses, are related to the Clebsch-Gordan
coefficients.  They are widely tabulated and easily computed for
applications (Appendix A).

In terms of these functions, the Hamiltonian for the atom-field
interaction is
\begin{equation}
\label{atom_Stark_Hamiltonian}
 H_{\rm el} = -(-e{\vec r}) \cdot
{\vec {\cal E}} = ez{\cal E} = er \cos (\theta){\cal E}  = er{\cal
E} C_{10}(\theta).
\end{equation}
The possible energies for a dipole in a field are given by the
eigenvalues of the Hamiltonian (\ref{atom_Stark_Hamiltonian}).  To
evaluate these energies in quantum mechanics, we identify the usual
basis set of hydrogenic wave functions, $|nlm \rangle$, where we
ignore spin for this simple illustration:
\begin{eqnarray}
\langle r,\theta,\phi | nlm \rangle = f_{nl}(r)Y_{lm}(\theta, \phi).
\nonumber
\end{eqnarray}
The matrix elements between any two hydrogenic states are
\begin{eqnarray}
\label{atom_integral} \langle nlm | -{\vec \mu} \cdot {\vec {\cal
E}} | n^{\prime}l^{\prime}m^{\prime} \rangle &&= \langle er \rangle
{\cal E} \int d(\cos \theta) d \phi Y^{*}_{lm} C_{10}
Y_{l^{\prime}m^{\prime}}
\\ &&= \langle er \rangle {\cal E} \sqrt{(2l+1)(2l^{\prime}+1)}
\left( \begin{array}{ccc} l & 1 & l^{\prime} \\
0 & 0 & 0 \end{array} \right)
\left( \begin{array}{ccc} l & 1 & l^{\prime} \\
-m & 0 & m^{\prime} \end{array} \right) \nonumber
\end{eqnarray}
Here $\langle er \rangle = \int r^2 dr f_{nl}(r) r
f_{n^{\prime}l^{\prime}}$ is an effective magnitude of the dipole
moment, which can be analytically evaluated for hydrogen
\cite{Bethe_Salpeter}.

Some important physics is embodied in Eqn. (\ref{atom_integral}).
First, the electric field defines an axis of rotational symmetry
(here the $z$ axis).  On general grounds, we therefore expect that
the projection of the total angular momentum of the molecule onto
this axis is a constant. And indeed, this is built into the 3-$j$
symbols: since the sum of all $m$ quantum numbers in a 3-$j$ symbol
should add to zero, Eqn. (\ref{atom_integral}) asserts that
$m=m^{\prime}$, and the electric field cannot couple two different
$m$'s together.

A second feature embodied in (\ref{atom_integral}) is the action of
parity.  The hydrogenic wavefunctions have a definite parity, i.e.,
they either change sign, or else remain invariant, upon converting
from a coordinate system $(x,y,z)$ to a coordinate system
$(-x,-y,-z)$. The sign of the parity-changed wave function is given
by $(-1)^l$. Thus an $s$-state ($l=0$) has even parity, while a
$p$-state ($l=1$) has odd parity. For an electric field pointing in
a particular direction, the Hamiltonian
(\ref{atom_Stark_Hamiltonian}) itself has odd parity, and thus
serves to {\it change} the parity of the atom. For example, it can
couple the  $s$ and $p$ states to each other, but not to themselves.
This is expressed in (\ref{atom_integral}) by the first 3-$j$
symbol, whose symmetry properties require that $l+1+l^{'}={\rm
even}$.  This seemingly innocuous statement is the {\it fundamental
fact} of electric dipole moments of atoms and molecules. It says
that, for example the $1s$ ground state of hydrogen, with
$l=l^{\prime}=0$, does not, by itself, respond to an electric field
at all.  Rather, it requires an admixture of a $p$ state to develop
a dipole moment.\footnote{These remarks are not strictly true.  The
ground state of hydrogen already has a small admixture of odd-parity
states, due to the parity-violating part of the electroweak force.
This effect is far too small to concern us here, however.}

To evaluate the influence of an electric field on hydrogen,
therefore, we must consider at least the nearest state of opposite
parity, which is the $2p$ state.  These two states are separated in
energy by an amount $E_{1s2p}$.  Considering only these two states,
and ignoring any spin structure, the atom-plus-field Hamiltonian is
represented by a simple $2\times2$ matrix:
\begin{equation}\label{atom_matrix}
H = \left( \begin{array}{cc} -E_{1s2p}/2 & \mu {\cal E} \\
\mu {\cal E} & E_{1s2p}/2 \end{array} \right).
\end{equation}
where the dipole matrix element is given by the convenient shorthand
$\mu = \langle 1s,m=0 | ez | 2p,m=0 \rangle = 128\sqrt{2} ea_0 / 243$
\cite{Bethe_Salpeter}. Of course there are many more $p$ states that
the $1s$ state is coupled to. Plus, all states are further
complicated by the spin of the electron and (in hydrogen) the
nucleus.  Matrix elements for all these can be constructed, and the
full matrix diagonalized to approximate the energies to any desired
degree of accuracy. However, we are interested here in the
qualitative features of dipoles, and so limit ourselves to
Eqf.(\ref{atom_matrix}).

The Stark energies are thus given approximately  by
\begin{eqnarray}
\label{atom_energies} E_{\pm} = \pm \sqrt{ (d{\cal E})^2 +
(E_{1s2p}/2)^2}.
\end{eqnarray}
This expression illustrates the basic physics of the quantum
mechanical dipole.  First, there are necessarily two states (or
more) involved.  One state decreases in energy as the field is
turned on, representing the``normal'' case where the electron moves
to negative $z$ and the electric dipole moment aligns with the
field. The other state, however, increases in energy with increasing
field and represents the dipole moment anti-aligning with the field.
Classically it is of course possible to align the dipole against the
field in a state of unstable equilibrium. Similarly, in quantum
mechanics this is a legitimate energy eigenstate, and the dipole
will remain anti-aligned with the field in the absence of
perturbations.

A second observation about the energies (\ref{atom_energies}) is
that the energy is a quadratic function of ${\cal E}$ at low field,
and only becomes linearly proportional to ${\cal E}$ at higher
fields.  Thus the permanent dipole moment of the atom, defined by
the zero-field limit
\begin{eqnarray}
\mu_{\rm  permanent} \equiv \lim_{{\cal E} \rightarrow 0} {
\partial E_- \over \partial {\cal E}}, \nonumber
\end{eqnarray}
vanishes.  The atom, in an energy eigenstate in zero field,
has no permanent electric dipole moment.  This
makes sense since, in zero field, the electron's position is
randomly varied about the atom, lying as much on one side of the
nucleus as on the opposite side.

The transition from quadratic to linear Stark effect is an example
of a competition between two tendencies.  At low field, the dominant
energy scale is the energy splitting $E_{1s2p}$ between opposite
parity states. At higher field values, the interaction energy with
the electric field becomes stronger, and the dipole is aligned.  The
value of the field where this transition occurs is found roughly by
setting these energies equal to find a ``critical'' electric field:
\begin{eqnarray}
{\cal E}_{\rm critical} = E_{1s2p}/2d. \nonumber
\end{eqnarray}
For atomic hydrogen, this field is on the order of $10^9$ V/cm.
However, at this field it is already a bad approximation to ignore
that fact that there are both $2p_{1/2}$ and $2p_{3/2}$ states, as
well as higher-lying $p$ states, and further coupling between $p$,
$d$, etc., states.  We will not pursue this subject further here.

\subsection{Rotating molecules}

With these basics in mind, we can move on to molecules.  We focus
here on diatomic, heteronuclear molecules, although the principles
 are more general.  We will consider only electric fields so small
that the electrons cannot be polarized in the sense of the previous
section; thus we consider only a single electronic state. However,
the charge separation between the two atoms produces an electric
dipole moment ${\vec \mu}$ in the rotating frame of the molecule.
We assume that the molecule is a rigid rotor and we
will not consider explicitly the vibrational motion of the molecule,
focusing instead solely on the molecular
rotation. (More precisely, we consider ${\vec \mu}$ to incorporate an
averaging over the vibrational coordinate of the molecule, much as
the electron-proton distance $r$ was averaged over for the hydrogen
atom in the previous section.)

As a mathematical preliminary, we note the following.  To deal with
molecules, we are required to transform freely between the
laboratory reference frame and the body-fixed frame that rotates
with the molecule.  The rotation from the lab frame $(x,y,z)$ to the
body frame $(x^{\prime},y^{\prime},z^{\prime})$ is governed by a set
of Euler angles $(\alpha, \beta, \gamma)$ (Appendix A). The first two angles
$\alpha = \phi$, $\beta = \theta$ coincide with the spherical
coordinates $(\theta, \phi)$ of the body frame's $z^{\prime}$ axis.
By convention, we take the positive $z^{\prime}$ direction to be
parallel to the dipole moment ${\vec \mu}$. The third Euler angle
$\gamma$ serves to orient the $x^{\prime}$ axis in a desired
orientation within the body frame; it is thus the azimuthal angle of
rotation about the molecular axis.

Consider a given angular momentum state $|jm \rangle$ referred to
the lab frame.  This state is only a state of good $m$ in the lab
frame, in general.  In the body frame, which points in some other
direction, the same state will be a linear superposition of
different $m$'s, which we denote in the body frame as $\omega$'s to
distinguish them.  Moreover, this linear superposition will be a
function of the Euler angles, with a transformation that is
conventionally denoted by the letter $D$:
\begin{eqnarray}
D(\alpha \beta \gamma)|j \omega \rangle &&= \sum_m |jm \rangle
\langle jm | D(\alpha, \beta,
\gamma) | j \omega \rangle \nonumber \\
&& \equiv \sum_m |jm \rangle D^j_{m\omega}(\alpha, \beta, \gamma).
\nonumber
\end{eqnarray}
This last line defines the Wigner rotation matrices, whose
properties are widely tabulated.  For each $j$, $D^j_{m\omega}$ is a
unitary transformation matrix; note that a rotation can only change
$m$-type quantum numbers, not the total angular momentum $j$.  One
of the more useful properties of the $D$ matrices, for us, is
\begin{eqnarray}
\label{D_integral} \int d\alpha d \cos(\beta) d\gamma &&
D^{j_1}_{m_1\omega_1}(\alpha \beta \gamma)
D^{j_2}_{m_2\omega_2}(\alpha \beta
\gamma)D^{j_3}_{m_3\omega_3}(\alpha \beta \gamma) = \nonumber \\ &&
8 \pi^2 \left(
\begin{array}{ccc} j_1 & j_2 & j_3 \\ m_1 & m_2 & m_3 \end{array}
\right) \left(
\begin{array}{ccc} j_1 & j_2 & j_3 \\ \omega_1 & \omega_2 & m\omega_3 \end{array}
\right)
\end{eqnarray}

Because the dipole is aligned along the molecular axis, and because
the molecular axis is tilted at an angle $\beta$ with respect to the
field, and because the field defines the $z$-axis,  the dipole
moment is defined by its magnitude $\mu$ times a unit vector with
polar coordinates $(\beta, \alpha)$.  The Hamiltonian for the
molecule-field interaction is given by
\begin{equation}
\label{rotating_Hamiltonian} H_{\rm el} = -{\vec \mu}_{\rm el} \cdot
{\vec {\cal E}} = -\mu {\cal E} C_{10}(\beta \alpha) = -\mu {\cal E}
D^{j*}_{q0}(\alpha \beta \gamma).
\end{equation}
For use below, we have taken the liberty of rewriting $C_{10}$ as a
$D$-function; since the second index of $D$ is zero, this function
does not actually depend on $\gamma$, so introducing this variable
is not as drastic as it seems.

To evaluate energies in quantum mechanics we need to choose a basis
set and take matrix elements. The Wigner rotation matrices are the
quantum mechanical eigenfunctions of the rigid rotor.  With
normalization, these wave functions are
\begin{eqnarray}
\langle \alpha \beta \gamma | nm_n \lambda_n \rangle = \sqrt{ { 2n+1
\over 8 \pi^2} } D^{n*}_{m_n \lambda_n}. \nonumber
\end{eqnarray}
As we did for hydrogen, we here ignore spin.  Thus $n$ is the
quantum number of rotation of the atoms about their center of mass,
$m_n$ is the projection of this angular momentum in the lab frame,
and $\lambda_n$ is its projection in the body frame.  In this basis,
the matrix elements of the Stark interaction are computed using
(\ref{D_integral}), to yield
\begin{eqnarray}
\label{rotating_matrix_element} && \langle nm_n \lambda_n | -{\vec
\mu}_{\rm el} \cdot {\vec {\cal
E}} | n^{'} m_n^{'} \lambda^{'} \rangle \\
&& = -\mu_{\rm el} {\cal E} (-1)^{m_n - \lambda_n}
\sqrt{(2n+1)(2n^{'}+1)} \left( \begin{array}{ccc} n & 1 & n ^{'} \\
-m_n & 0 & m_n^{'} \end{array} \right) \left( \begin{array}{ccc} n & 1 & n ^{'} \\
-\lambda_n & 0 & \lambda_n^{'} \end{array} \right). \nonumber
\end{eqnarray}
In an important special case, the molecule is in a $\Sigma$ state,
meaning that the electronic angular momentum projection
$\lambda_n=0$.  In this case, Eqn. (\ref{rotating_matrix_element})
reduces to the same expression as that for hydrogen, apart from the
radial integral.  This is as it should be: in both objects, there is
simply a positive charge at one end and a negative charge at the
other.  It does not matter if one of these is an electron, rather
than an atom.  More generally, however, when $\lambda_n \ne 0$ there
will be a complicating effect of lambda-doubling, which we will
discuss in the next section.

Thus the physics of the rotating dipole is much the same as that of
the hydrogen atom.  Eqn. (\ref{rotating_matrix_element}) also
asserts that, for a $\Sigma$ state with $\lambda_n=0$, the electric
field interaction vanishes unless $n$ and $n^{\prime}$ have opposite
parity. For such a state, the parity is related to the parity of $n$
itself. Thus, for the ground state of a $^1\Sigma$ molecule with
$n=0$, the electric field only has an effect by mixing this state
with the next rotational state with $n^{\prime} = 1$.  These states
are split by an energy $E_{\rm rot} = 2B_e$, where $B_e$ is the
rotational constant of the molecule.\footnote{In zero field, the
state with rotational quantum number $n$ has energy $B_en(n+1)$.}

We can formulate a simple $2 \times 2$ matrix describing this
situation, as we did for hydrogen:
\begin{equation}\label{rotating_matrix} H = \left( \begin{array}{cc}
-E_{\rm rot}/2 & -\mu {\cal E} \\ -\mu {\cal E} & +E_{\rm rot}/2
\end{array}\right),
\end{equation}
where the dipole matrix element is given by the convenient shorthand
notation
$\mu = \langle n m_n 0 | \mu_{q=0} | n^{\prime} m_n 0 \rangle$.
There is one such matrix for each value of $m_n$.  Of course there
are many more rotational states that these states are coupled to.
Plus, all states would further be complicated by the spins (if any)
of the electrons and nuclei.

The matrix (\ref{rotating_matrix}) can be diagonalized just as
(\ref{atom_matrix}) was above, and the same physical conclusions
apply.  Namely, the molecule in a given rotational state
has {\it no permanent electric dipole
moment}, even though there is a separation of charges in the body
frame of the molecule.  Second, the Stark effect is quadratic for
low fields, and linear only at higher fields, with the transition
occurring at a ``critical field''
\begin{equation}
E_{\rm crit} = E_{\rm rot}/2\mu.
\end{equation}
To take an example, the NH molecule posesses a $^3 \Sigma$ ground
state.  For this state, ignoring spin, the critical field is of the
order $7 \times 10^{6}$ V/cm. This is far smaller than the field
required to polarize electrons in an atom or molecule, but still
large for laboratory-strength electric fields.  Diatomic molecules
with smaller rotational constants, such as LiF, would have correspondingly
smaller critical fields. In any event, by the time the critical field
is applied,  it is already a bad approximation to ignore coupling to
the other rotational states of the molecule, which  must be included
for an accurate treatment. We do not consider this topic further
here.

\subsection{Molecules with lambda-doubling}

As we have made clear in the previous two sections, the effect of an
electric field on a quantum mechanical object is to couple states of
opposite parity.  For a molecule in a $\Pi$ or $\Delta$ state, there
are often two such parity states that are much closer together in
energy than the rotational spacing.  The two states are said to be
the components of a ``$\Lambda$-doublet.'' Because they are close
together in energy, these two states can then be mixed at much
smaller fields than are required to mix rotational levels.  The
physics underlying the lambda doublet is rather complex, and we
refer the reader to the literature for details \cite{BC,Hougen}.

However, in broad terms, the argument is something like this: a
$\Pi$ state has an electronic angular momentum projection of
magnitude 1 about the molecular axis. This angular momentum comes in
two projections, for the two sense of rotation about the axis, and
these projections are nominally degenerate in energy.  The rotation
of the molecule, however, can break the degeneracy between these
levels, and (it so happens) the resulting nondegenerate
eigenfunctions are also eigenfunctions of parity.  The main point is
that the resulting energy splitting is usually quite small, and
these parity states can be mixed in fields much smaller than those
required to mix rotational states.

To this end, we modify the rigid-rotor wave function of the molecule
to incorporate the electronic angular momentum:
\begin{equation}
\label{doublet_basis} \langle \alpha \beta \gamma | j m \omega
\rangle = \sqrt{ {2j+1 \over 8\pi^2} } D^{j*}_{m \omega} (\alpha
\beta \gamma).
\end{equation}
Here $j$ is the {\it total} (rotation-plus-electronic) angular
momentum of the molecule, and $m$ and $\omega$ are the projections
of $j$ on the laboratory and body-fixed axes, respectively.  Using
the total $j$ angular momentum, rather than just the molecular
rotation $n$, marks the use of a ``Hund's case a'' representation,
rather than the Hund's case b that was implicit in the previous
section (Ref. \cite{BC}; see also Appendix B of this volume).

In this basis the matrix element of the electric field Hamiltonian
(\ref{rotating_Hamiltonian}) becomes
\begin{eqnarray}
\label{doublet_matrix_element} && \langle \j m \omega | -{\vec
\mu}_{\rm el} \cdot {\vec {\cal E}} | j^{\prime} m^{\prime} \omega
^{\prime} \rangle
\\ && = - \mu_{\rm el} {\cal
E} (-1)^{m-\omega} (2j+1) \left( \begin{array}{ccc} j & 1 & j^{\prime} \\
-m & 0 & m^{\prime} \end{array} \right) \left( \begin{array}{ccc} j
& 1 & j^{\prime} \\ -\omega & 0 & \omega^{\prime} \end{array}
\right). \nonumber
\end{eqnarray}
In (\ref{doublet_matrix_element}), the 3-$j$ symbols denote
conservation laws.  The first asserts that $m = m^{\prime}$ is
conserved, as we already knew.  The second 3-$j$ symbol adds to this
the fact that $\omega = \omega^{\prime}$.  This is the statement
that the electric field cannot exert a torque around the axis of the
dipole moment itself.  Moreover, in the present model we assert that
$j = j^{\prime}$, since the next higher-lying $j$ level is far away
in energy, and only weakly mixes with the ground state $j$. With
these approximations, the 3-$j$ symbols have simple algebraic
expressions, and we can simplify the matrix element:
\begin{eqnarray}
\label{doubled_Stark_energies} \langle j m \omega | -{\vec \mu}_{\rm
el} \cdot {\vec {\cal E}} | j m \omega \rangle = - \mu {\cal E} { m
\omega \over j(j+1)}. \nonumber
\end{eqnarray}
The physical content of this expression is illustrated in Fig. 1.
Notice that both $m$ and $\omega$ can have a sign, and that whether
the energy is positive or negative depends on both signs.

An essential point is that there is not a unique state representing
the dipole aligned with the field.  Rather, there are two such
states, distinguished by different angular momentum quantum numbers
but possessing the same energy.  To distinguish these in the
following we will refer to the two states in figures 1(a) and 1(b)
as molecules of type $|a \rangle$ and type $|b \rangle$,
respectively. Likewise, for molecules nominally anti-aligned with
the field, we will refer to types $|c \rangle$ and $|d \rangle$,
corresponding to the two states in figures 1(c) and 1(d).  The
existence of these degeneracies will lead to novel phenomena in
these kinds of molecules, as we discuss below.

As to the lambda doubling, it is, as we have asserted, diagonal in a
basis where parity is a good quantum number.  In terms of the basis
(\ref{doublet_basis}), wave functions of well-defined parity are given by
the linear combinations
\begin{equation}
\label{parity_basis} |j m {\bar \omega } \epsilon \rangle = {1 \over
\sqrt{2} } \left[ |j m {\bar \omega} \rangle + \epsilon |j m {\bar
\omega} \rangle \right].
\end{equation}
Here we define ${\bar \omega} = |\omega|$, the absolute value of
$\omega$.    For a given value of $m$, the linear combinations of
$\pm {\bar \omega}$ in (\ref{parity_basis}) are distinguished by the
parity quantum number $\epsilon = \pm 1$. It is straightforward to show
that in the parity basis (\ref{parity_basis}) the $\Lambda$-doubling
is off-diagonal.

\begin{figure}
\caption{Energetics of a polar molecule in an electric field.  The
molecule's dipole moment points from the negatively charged atom
(large circle) to the positively charged atom (small circle) as
indicated by the thick arrow.  The dashed line indicates the
positive direction of the molecules body axis, while the vertical
arrow represents the direction of the applied electric field.  The
dipole aligns with the field, on average, if {\it either} i) the
angular momentum ${\vec j}$ aligns with the field and $m>0$, $\omega
> 0$ (a); or ii) ${\vec j}$ aligns against the field and $m<0$,
$\omega <0$ (b). Similar remarks apply to dipoles that anti-align
with the field (c,d).} \centering
\includegraphics{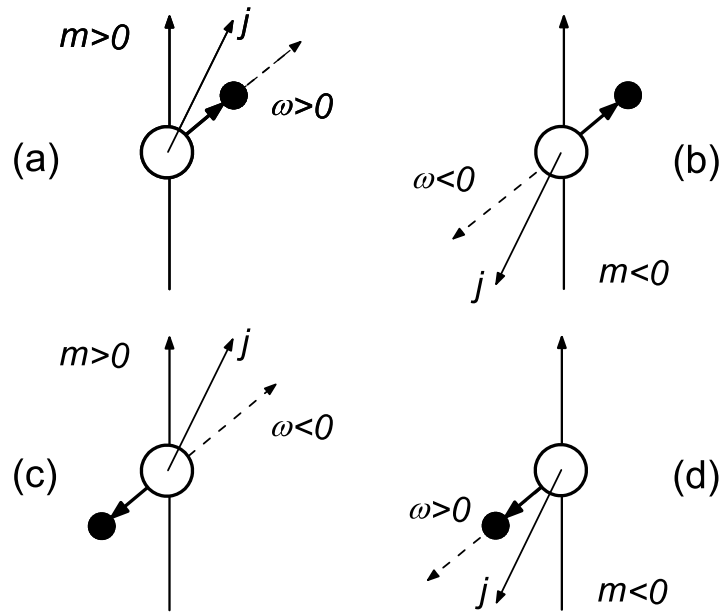}
\end{figure}

The net result is that for each value of $m$, the  Hamiltonian for
our lambda-doubled molecule can be represented as a two-by-two
matrix, similar to the ones above:
\begin{eqnarray}
\label{doublet_matrix}  H = \left( \begin{array}{cc} -Q & \Delta/2 \\
\Delta/2 & Q
\end{array} \right), \nonumber
\end{eqnarray}
where $\Delta$ is the lambda doubling energy, i.e., the energy
difference between the two parity states, and
\begin{eqnarray}
Q \equiv  \mu {\cal E} { | m | {\bar \omega} \over j(j+1) }
\nonumber
\end{eqnarray}
is a manifestly positive quantity.  This is the Hamiltonian we
will treat in the remainder of this chapter.  The difference here
from the previous subsections is that the basis is now
(\ref{doublet_basis}), which diagonalizes the electric field
interaction rather than the zero-field Hamiltonian.  This change
reflects our emphasis on molecules in strong fields where their
dipole moments are made manifest.  The zero-field $\Lambda$-doubling
Hamiltonian is considered, for the most part, to be a perturbation.

The mixing of the strong field states due to the $\Lambda$-doubling
interaction is conveniently given by a mixing angle $\delta_m$,
which we define as follows:
\begin{eqnarray}
\label{eigenstates} | m {\bar \omega} \epsilon=+ \rangle &=& \cos
\delta_m |m {\bar \omega} \rangle + \sin \delta_m | m -{\bar \omega}
\rangle
\nonumber \\
|m {\bar \omega} \epsilon=- \rangle &=& -\sin \delta_m | m {\bar
\omega} \rangle + \cos \delta_m | m -{\bar \omega} \rangle
\end{eqnarray}
Explicitly, the mixing angle as a function of field is given by
\begin{eqnarray}
\tan \delta_{|m|} = {\Delta/2 \over Q + Q\sqrt{1 + \eta^2} } = -\tan
\delta_{-|m|}, \nonumber
\end{eqnarray}
in terms of the energy $Q$ and the dimensionless parameter
\begin{eqnarray}
\eta_m = { \Delta \over 2Q}. \nonumber
\end{eqnarray}
Notice that with this definition, $\delta_m$ is positive when $m$ is
positive, and $\delta_{-m} = -\delta_m$.  The energies of these
states are conveniently summarized by the expression
\begin{eqnarray}
\label{eigenenergies} E_{m{\bar \omega} \epsilon} = -\mu {\cal E} {
m\epsilon {\bar \omega} \over j(j+1)}\sqrt{ 1+ \eta^2}, \;\;\;\;\; m
\ne 0. \nonumber
\end{eqnarray}
This is a very compact way of writing the results that will
facilitate writing the expressions below.  Notice that the intuition
afforded by Figure 1 is still intact in these energies, but by
replacing the sign of $\epsilon$ for the sign of $\omega$.  Thus
states with $m \epsilon > 0$ have negative energy, while states with
$m \epsilon <0$ have positive energy.

The case where  $m=0$ must be handled slightly differently, since in
this case the electric field energies $Q=0$. We can still write the
eigenstates in the form (\ref{eigenstates}), provided that we set
\begin{eqnarray}
\delta_{0} = {\pi \over 4}, \nonumber
\end{eqnarray}
and understand that the corresponding energies, independent of
field, are simply
\begin{eqnarray}
E_{0{\bar \omega}\epsilon} = - \epsilon {\Delta / 2}. \nonumber
\end{eqnarray}

Like the dipoles considered above, this model has a quadratic Stark
effect at low energies, rolling over to linear at electric fields
exceeding the critical field given by setting $Q = \Delta/2$.  This
criterion gives a critical field of
\begin{equation}
E_{\rm crit} = {\Delta j(j+1) \over 2 \mu |m| {\bar \omega}}.
\end{equation}
To take an example, consider the ground state of OH, which has $j=
{\bar \omega} =3/2$, $\mu = 1.7$ Db, and a $\Lambda$-doublet
splitting of $0.06$ cm$^{-1}$.  In its $m=3/2$ ground state, its
critical field is $\sim 1600$ V/cm. Again, we have explicitly
ignored the spin of the electron.  The parity states are therefore
easily mixed in fields that are both small enough to easily obtain
in the laboratory, and small enough that no second-order coupling to
rotational or electronic states needs to be considered.  Keeping the
relatively small number of molecular states is therefore a
reasonable approximation, and highly desirable as it simplifies our
discussion.\footnote{In OH there is also the ${\bar \omega} = 1/2$
state to consider, but it is also far away in energy as compared to
the $\Lambda$-doublet, and we ignore it. It does play a role in the
fine structure of OH, however, and this should be included in a
quantitative model of OH.}


\section{The field due to a dipole}

Once polarized, each of the $\Lambda$-doubled molecules discussed
above is itself the source of an electric field.  The field due to a
molecule is given above by Eqn. (\ref{classical_dipole}).  In what
follows, it is convenient to cast this potential in terms of the
spherical tensors defined above and in Appendix A:
\begin{eqnarray}
\label{dipole_potential} \Phi ({\vec r}) = { {\vec \mu} \cdot {\hat
r} \over r^2} = {\mu \over r^2} \sum_q (-1)^q C_{1q}( \alpha \beta)
C_{1-q}(\theta \phi),
\end{eqnarray}
That this form is correct can be verified by simple substitution,
using the definitions (\ref{C_functions}). It seems at first
unnecessarily complicated to write (\ref{dipole_potential}) in this
way. However, the effort required to do so will be rewarded when we
need to evaluate the potential for quantum mechanical dipoles below.

For a classical dipole, defining the direction $(\alpha \beta)$ of
its dipole moment would immediately specify the electrostatic
potential it generates according to Eq. (\ref{dipole_potential}).
However, in quantum mechanics the potnetial will result from
suitably averaging the orientation of $\mu$ over the distribution of
$(\alpha \beta)$ dictated by the molecule's wave function.  To
evaluate this, we need to evaluate matrix elements of
(\ref{dipole_potential}) in the basis of energy eigenstates
(\ref{eigenstates}).  This is easily done using
Eq.(\ref{D_integral}), along with formulas that simplify the 3-$j$
symbols.  The result is
\begin{eqnarray}\label{C_matrix_elements}
\langle j m \omega | C_{1q} | j m^{\prime} \omega^{\prime} \rangle
 = \delta_{\omega \omega^{\prime}} {\omega \over j(j+1) }  \left\{ \begin{array}{ll} -\sqrt{
{(j+m)(j-m^{'}) \over 2}}, & q=+1 \\ m, & q=0 \\ +\sqrt{
{(j-m)(j+m^{'}) \over 2}}, & q=-1 \end{array} \right\},
\end{eqnarray}
where the quantum numbers $m$ and $\omega$ are signed quantities.
Also note that this integration is over the molecular degrees of
freedom $(\alpha \beta)$ in (\ref{dipole_potential}).  This still
leaves the angular dependence on $(\theta \phi)$, which
characterizes the field in space around the dipole.

From expression (\ref{C_matrix_elements}) it is clear that this
matrix element changes sign upon either i) reversing the sign of
both $m$ and $m^{\prime}$ or ii) changing the sign of $\omega$.
Moreover, $\langle j m \omega | C_{1q} | j m^{\prime} -\omega
\rangle = 0$, since $\omega$ is conserved by the electric field.
Using these observations, we can readily compute the matrix elements of
 $\Phi$ in the dressed basis (\ref{parity_basis}).  Generally they take the form
\begin{eqnarray} \label{dipole_matrix_elements}
\langle j m {\bar \omega} \epsilon | \Phi( {\vec r}) | j m^{\prime}
{\bar \omega} \epsilon^{\prime} \rangle = \langle j m {\bar \omega}
\epsilon | C_{1q} | j m^{\prime} {\bar \omega} \epsilon^{\prime}
\rangle (-1)^q {\mu \over r^2} C_{1-q}(\theta \phi).
\end{eqnarray}
The matrix elements in front of (\ref{dipole_matrix_elements})
represent the quantum mechanical manifestation of the dipole's
orientation.  These matrix elements follow from the above definition
of the dressed states, (\ref{eigenstates}).  Explicitly,
 \begin{eqnarray}
 \label{C_matrix_elements_dressed}
 \langle j m {\bar \omega}, \epsilon | C_{1q} | j m^{\prime} {\bar
 \omega}, \epsilon \rangle &=& \epsilon  \cos (\delta_m + \delta_{m^{\prime}}) \langle j m {\bar
 \omega}
 | C_{1q} |j m^{\prime} {\bar \omega} \rangle
 \nonumber \\
  \langle j m {\bar \omega},- | C_{1q} | j m^{\prime} {\bar
 \omega}, + \rangle &=& \langle j m {\bar \omega},+ | C_{1q} | j m^{\prime} {\bar
 \omega}, - \rangle  \\ &=&  -\sin (\delta_m + \delta_{m^{\prime}})  \langle j m {\bar
 \omega}
 | C_{1q} |j m^{\prime} {\bar \omega} \rangle .
 \nonumber
 \end{eqnarray}
 In all these expressions, the value of $q$ is set by angular
 momentum conservation to $q = m - m^{\prime}$.

This description, while complete within our model, nevertheless
remains somewhat opaque.  Let us therefore specialize it to the case of
a particular energy eigenstate $|jm{\bar \omega} \epsilon \rangle$.
In this state, the averaged electrostatic potential of the dipole is
\begin{equation} \label{aligned_dipole}
\langle \Phi ({\vec r}) \rangle = \left( \mu {m \epsilon \omega
\over j(j+1)} \cos 2\delta_m \right) {\cos \theta \over r^2}.
\end{equation}
Here the factor in parentheses is a quantum mechanical correction to
the magnitude of the dipole moment.  The factor $\cos 2\delta_m$,
expresses the degree of polarization: in a strong field, $\delta_m =
0$ and the dipole is at maximum strength, whereas in zero field
$\delta_m= \pi/4$ and the dipole vanishes. Notice that for a given
value of $m$, the potential generated by the states with $\epsilon =
\pm$ differ by a sign.  this is appropriate, since these states
correspond to dipoles pointing in opposite directions (Figure 1).

The off-diagonal matrix elements in (\ref{dipole_matrix_elements})
are also important, for two reasons.  First, it may be desirable to
create superpositions of different energy eigenstates, and computing
these matrix elements requires the off-diagonal elements in
(\ref{dipole_matrix_elements}), as we will see shortly.  Second,
when two dipoles interact with each other, one will experience the
electric field due to the other, and this field need not lie
parallel to the $z$-axis.  Hence, the $m$ quantum number of an
individual dipole is no longer conserved, and elements of
({\ref{dipole_matrix_elements}) with $q \ne 0$ are required.

\subsection{Example: $j=1/2$}

To illustrate these abstract points, we consider here the simplest
molecular state with a $\Lambda$ doublet:  a molecule with
$j=1/2$, which has ${\bar \omega} = 1/2$, and consists of four
internal states in our model.  Based on the discussion above, we
tabulate the matrix elements between these states in Table I.
Because we have $j=1/2$, we suppress the index $j$ in this section.

For concreteness, we focus on a type $|a \rangle$  molecule, as
defined in Figure 1.  This molecule aligns with the field and
produces an electrostatic potential (\ref{aligned_dipole}). However,
if the molecule is prepared in a state that is a superposition of
this state with another, a different electrostatic potential can
result.  We first note that combining $|a \rangle$ with $|b
\rangle$ produces nothing new, since both states generate the same
potential.

An alternative superposition combines states $|a \rangle$ with state
$|c \rangle$.  In this case the two states have the same value of
$m$, but are nevertheless non-degenerate.  We define
\begin{eqnarray}
\label{ac} |\psi \rangle_{ac} = A e^{i \omega_0 t} | {1 \over 2}
{\bar \omega},+ \rangle + B e^{-i\omega_0 t} | {1 \over 2} {\bar
\omega}, - \rangle, \nonumber
\end{eqnarray}
for arbitrary complex numbers $A$ and $B$ with $|A|^2+|B|^2=1$.
Because the two states are non-degenerate, it is necessary to
include the explicit time-dependent phase factors, where $\omega_{0}
= |E_{m {\bar \omega} \epsilon}|/\hbar$, and $2\hbar \omega_0$ is
the energy difference between the states. As usual in quantum
mechanics, these phases will beat against one another to make the
observables time dependent.

\begin{table}
\caption{Matrix elements $\langle m {\bar \omega} \epsilon | C_{1q}
| m^{\prime} {\bar \omega} \epsilon^{\prime} \rangle$ for a $j=1/2$
molecule. To obtain matrix elements of the electrostatic potential
$\Phi({\vec r})$, these matrix elements should be multiplied by $
(-1)^q C_{k-q}(\theta\phi)\mu/r^2$, where $q=m-m^{\prime}$. } \vskip
0.2in \center
\begin{tabular}{r | c c c c}
 & $|{1 \over 2} {\bar \omega}+ \rangle$ & $|{1 \over 2} {\bar \omega}- \rangle$
  & $|-{1 \over 2} {\bar \omega}+ \rangle$ & $|-{1 \over 2} {\bar \omega}- \rangle$ \\
 \hline
 $\langle {1 \over 2} {\bar \omega}+ |$ & $ {1 \over 3} \cos 2\delta_{1/2} $ &
 $-{1 \over 3} \sin 2\delta_{1/2} $ & $-{\sqrt{2} \over 3}$ & $0$ \\
$\langle {1 \over 2} {\bar \omega}- |$ & $-{1 \over 3} \sin
2\delta_{1/2}$ &
$-{1 \over 3} \cos 2\delta_{1/2}$ & $0$ & ${\sqrt{2} \over 3}$ \\
$\langle -{1 \over 2} {\bar \omega}+ |$ & ${\sqrt{2} \over 3}$ & $0$
&
 $-{1 \over 3} \cos 2\delta_{1/2}$ & $-{1 \over 3} \sin 2\delta_{1/2}$  \\
$\langle -{1 \over 2} {\bar \omega}- |$ & $0$ & $-{\sqrt{2} \over
3}$ & $ -{1 \over 3} \sin 2\delta_{1/2}$ & ${1 \over 3} \cos
2\delta_{1/2}$
 \end{tabular}
\end{table}

Now some algebra identifies the mean value of the electrostatic
potential, averaged over state $|\psi \rangle_{ac}$, as
\begin{eqnarray} \label{bobbing_dipole}
&& _{ac} \langle \psi | \Phi ({\vec r}) | \psi \rangle_{ac}
\\ &&= {\mu  \over 3 r^2} \left[ \left( |A|^2 -
|B|^2 \right) \cos 2\delta_{1/2} - 2 |AB| \sin 2\delta_{1/2} \cos(2
\omega_0 t - \delta) \right] \cos \theta. \nonumber
\end{eqnarray}
This potential has the usual $\cos \theta$ angular dependence,
meaning that the dipole remains aligned along the field's axis.
However, the magnitude, and even the sign, of the dipole change over
time.  The first term in square brackets in (\ref{bobbing_dipole})
gives a constant, dc component to the dipole moment, which depends
on the population imbalance $|A|^2-|B|^2$ between the two states.
The second term adds to this an oscillating component with angular
frequency $2\omega_0$.  The leftover phase $\delta$ is an irrelevant
offset, and comes from the phase of $A^*B$, i.e., the relative phase
of the two components at time $t=0$.

It is therefore possible to construct a superposition of states of
the dipole, such that the effective dipole moment of the molecule
bobs up and down in time.  The amount that the dipole bobs, relative
to the constant component, can be controlled by the relative
population in the two states.  Moreover, the degree of polarization
of the molecule plays a significant role.  For a fully polarized
molecule, when $\delta_{1/2} = 0$, only the dc portion of the dipole
persists, although even it can vanish if there is equal population
in the two states, ``dipole up'' and ``dipole down.''

As another example, we consider the superposition of $|a \rangle$
with $|d \rangle$.  Now the two states have different values of $m$
as well as different energies:
\begin{eqnarray} \label{rotating}
| \psi \rangle_{ad} =  A e^{i \omega_0 t} | {1 \over 2} {\bar
\omega},+ \rangle + B e^{-i\omega_0 t} | -{1 \over 2} {\bar \omega},
+ \rangle. \nonumber
\end{eqnarray}
The dipole potential this superposition generates is
\begin{eqnarray} \label{rotating_dipole}
&& _{ad} \langle \psi | \Phi({\vec r}) | \psi \rangle_{ad} \nonumber
\\ &&= {\mu \over 3 r^2}  \cos 2 \delta_{1/2}
\left( |A|^2 - |B|^2 \right) \cos \theta +  {2 \mu \over 3 r^2} {\rm
Re} \left[ A^*B e^{-i(\phi + 2\omega_0 t)} \right] \sin \theta.
\nonumber
\end{eqnarray}
This expression can be put in a useful and interesting form if we
parametrize the coefficients $A$ and $B$ as
\begin{eqnarray} \label{AB_param}
A = \cos {\alpha \over 2} e^{-i \beta/2} \;\;\;\; B = \sin {\alpha
\over 2} e^{i \beta/2}. \nonumber
\end{eqnarray}
This way of writing $A$ and $B$ seems arbitrary, but it is not.  It
is the same parametrization that is used in constructing the Bloch
sphere, which is a powerful tool in the analysis of any two-level
system \cite{Bloch}.

This parametrization leads to the following expression for the
potential:
\begin{eqnarray} \label{rotating_dipoleII}
&& _{ad} \langle \psi | \Phi({\vec r}) | \psi \rangle_{ad}
\\ &&= {1 \over 3} {\mu \over r^2} \left[ \cos 2\delta_{1/2} \cos
\alpha \cos \theta + \sin \alpha \sin \theta \cos ( (\beta -
2\omega_0 t) - \phi) \right]. \nonumber
\end{eqnarray}
The interpretation of this result is clear upon comparing it to the
classical result (\ref{dot_product}).  First consider that the
molecule is perfectly polarized, so that $\cos 2\delta_{1/2} = 1$.
Then (\ref{rotating_dipoleII}) represents the potential due to a
dipole whose polar coordinates are $(\alpha, \beta - \omega t)$.
That is, this dipole makes (on average) an angle $\alpha$ with
respect to the field, and it precesses about the field with an
angular frequency $2\omega_0$.   Interestingly, even in this strong
field limit where the field nominally aligns the dipole along $z$,
quantum mechanics allows the dipole to point in quite a different
direction. As the field relaxes, the $z$-component reduces, but this
dipole still has a component precessing about the field.

\subsection{Example: $j=1$}

We also consider a molecule with spin $j=1$.  Here there are in
principle three mixing angles, $\delta_1$, $\delta_0$, and
$\delta_{-1}$.  However, as noted above we have $\delta_{-1} =
-\delta_1$ and $\delta_0 = \pi/4$, so that the entire electric field
dependence of these matrix elements is incorporated in the single
parameter $\delta_1$.   In this notation, the matrix elements of the
electrostatic potential for a $j=1$ molecule are given in Table 2.

Similar remarks apply to the spin-1 case as applied to the spin-1/2
case.  If the molecule is in an eigenstate, say $| +1 {\bar \omega},
- \rangle$, then the expectation value of the dipole points along
the field axis, and its distribution has the usual $\cos \theta$
dependence. In an eigenstate with $m=0$, however, the expectation
value of the dipole vanishes altogether.

As before, the molecule can also be in a superposition state.  No
matter how complicated this superposition is, the expectation value
of the dipole must instantaneously point in some direction, since
the only available angular dependence resides in the $C_{1q}$
functions, which yield only dipoles.  In other words, no
superposition can generate the field pattern of a quadrupole moment,
for example.

Where this dipole points, and how its orientation evolves with time,
however, can be complicated.  For example, a superposition of $|+1
{\bar \omega}, + \rangle$ and $|+1 {\bar \omega}, - \rangle$ can bob
up and down, just like the analogous superposition for $j=1/2$.
However, for $j=1$ molecules additional superpositions are
possible.  For example, consider the combination
\begin{eqnarray}
| \psi \rangle_{3} = A e^{i\omega_0 t} |+1 {\bar \omega}, + \rangle
+ B e^{i \omega_{\Delta}t} | 0 {\bar \omega}, + \rangle + C e^{i
\omega_0 t} | -1 {\bar \omega}, - \rangle, \nonumber
\end{eqnarray}
where $\hbar \omega_{\Delta} = \Delta/2$ is a shorthand notation
for half the lambda doubling energy.  Let us further assume for
convenience that $A$, $B$, and $C$ are all real. Then the
expectation value of the electrostatic potential is
\begin{eqnarray}
\label{sliding_dipole} _{3} \langle \psi | \Phi ({\vec r}) | \psi
\rangle_{3} &=& {\mu \over 2} \cos 2\delta_1 \left(
A^2 + C^2 \right) {\cos \theta \over r^2} \nonumber \\
&+& {\mu \over \sqrt{2} } \cos (\delta_1 + \pi /4) B {\sin \theta
\over r^2} \nonumber \\
& \times & \left[ A \cos((\omega_{\Delta} - \omega_0)t-\phi) -
C\cos( -(\omega_{\Delta} - \omega_0)t-\phi) \right]. \nonumber
\end{eqnarray}
By analogy with remarks in the previous section, this represents a
dipole with a constant component along the $z$-axis, which depends
on both the strength of the field and on $|A|^2 + |C|^2$, the total
population in the $\pm m$ states.  It also has a component in the
$x$-$y$ plane, orthogonal to the field's direction.  In the case
where $C=0$, this component would precess around the field axis with
a frequency $\omega_{\Delta} - \omega_0$, in a clockwise direction
as viewed from the $+z$ direction.  Vice-versa, if $A=0$, this
component would rotate at this frequency but in a counter-clockwise
direction.  If both components are present and $A=C$, then the
result will be, not a rotation, but an oscillation of this component
from, say, $+x$ to $-x$, in much the same way that linearly
polarized light in a superposition of left- and right-circularly
polarized components.  More generally, if $A \ne C$, then the tip of
the dipole moment will trace out an elliptical path.  However, in
the limit of zero field, $\omega_0$ reduces to $\omega_{\Delta}$ and
these time- dependent effects go away.

\begin{landscape}
\begin{table}
\caption{ Matrix elements of $C_{1q}$ for a $j=1$ molecule. To
obtain the matrix elements of the electrostatic potential $\Phi({\vec
r})$, these matrix elements should be multiplied by $(-1)^q
C_{k-q}(\theta\phi)\mu/r^2$, where $q=m-m^{\prime}$.}  \vskip 0.2in
\begin{tabular}{r | c c c c c c}
 & $|+1 {\bar \omega}+ \rangle$ & $| +1 {\bar \omega}- \rangle$ &
 $|0 {\bar \omega}+ \rangle$ & $|0 {\bar \omega}- \rangle$ & $|-1
 {\bar \omega}+ \rangle$ & $|-1 {\bar \omega}- \rangle$ \\
 \hline
 $\langle +1 {\bar \omega}+ |$ & ${1 \over 2} \cos 2\delta_1
 $ & $-{1 \over 2} \sin 2\delta_1 $ & $ -{1 \over 2} \cos
 (\delta_1 + \pi/4) $ & $ {1 \over 2} \sin (\delta_1 +
 \pi / 4) $ & 0 & 0 \\
 $\langle +1 {\bar \omega}- |$ & $-{1 \over 2} \sin 2\delta_1
 $ & $-{1 \over 2} \cos 2\delta_1 $ & ${1 \over 2} \sin
(\delta_1 + \pi /4) $ & ${1 \over 2} \cos ( \delta_1 + \pi
/4) $ & 0 & 0 \\
$\langle 0 {\bar \omega}+ |$ & ${1 \over 2} \cos (\delta_1 + \pi /4)
$ & $ -{1 \over 2} \sin (\delta_1 + \pi/4) $ & 0 & 0 & $-{1 \over 2}
\cos (-\delta_1 + \pi /4) $ & ${1 \over 2}
\sin (-\delta_1 + \pi /4) $ \\
$ \langle 0 {\bar \omega}- |$ & $-{1 \over 2} \sin (\delta_1 + \pi
/4) $ & $-{1 \over 2} \cos (\delta_1 + \pi /4) $ & 0 & 0 & ${1 \over
2} \sin (-\delta_1 + \pi /4) $ & ${1 \over 2} \cos
(-\delta_1 + \pi /4) $ \\
$\langle -1 {\bar \omega} + |$ & 0 & 0 & ${1 \over 2} \cos
(-\delta_1 + \pi /4) $ & $-{1 \over 2} \sin (-\delta_1 + \pi /4) $ &
$-{1 \over 2} \cos 2\delta_1 $ & $-{1 \over 2} \sin 2 \delta_1
$ \\
$\langle -1 {\bar \omega}- |$ & 0 & 0 & $-{1 \over 2} \sin
(-\delta_1 + \pi /4) $ & $-{1 \over 2} \cos (-\delta_1 + \pi /4) $ &
$-{1 \over 2} \sin 2\delta_1 $ & ${1 \over 2} \cos 2 \delta_1 $
\end{tabular}
\end{table}
\end{landscape}

\section{Interaction of dipoles}

Having thus carefully treated individual dipoles and their quantum
mechanical matrix elements, we are now in a position to do the same
for the dipole-dipole interaction between two molecules. This
interaction depends on the orientation of each dipole, ${\vec
\mu}_1$ and ${\vec \mu}_2$; and on their relative location, ${\vec
R}$.  This interaction has the form (Appendix A)
\begin{eqnarray}
\label{dd_interaction} V_d ({\vec r}) &=& { {\vec \mu}_1 \cdot {\vec
\mu}_2 - 3 (  {\vec \mu}_1 \cdot {\hat r}) (  {\vec \mu}_2 \cdot
{\hat r}) \over R^3 }  \\ &=& - {\sqrt{6} \mu^2 \over R^3} \sum_q
(-1)^q   \left[ \mu_1 \otimes \mu_2 \right]_{2q} C_{2-q}(\theta
\phi). \nonumber
\end{eqnarray}
In going from the first line to the second, we assume that both
molecules have the same size dipole moment $\mu$, and that the
intermolecular axis makes an angle $\theta$ with respect to the
laboratory $z$ axis, so that ${\vec R} = (R,\theta,\phi)$.  The
angles $\theta$ and $\phi$ thus stand for something slightly
different than in the previous section.  The third line in
(\ref{dd_interaction}) rewrites the interaction in a compact tensor
notation that is useful for the calculations we are about to do.
Here
\begin{eqnarray} \left[ \mu_1 \otimes\mu_2 \right]_{2q} = \sqrt{5} \sum_{q_1
q_2} (-1)^q \left( \begin{array}{ccc} 2 & 1 & 1 \\ q & -q_1 & -q_2
\end{array} \right) C_{1q_1}(\beta_1 \alpha_1) C_{1q_2}(\beta_2
\alpha_2) \nonumber
\end{eqnarray}
denotes the second-rank tensor composed of the two first-rank
tensors (i.e., vectors) $C_{1q_1}(\beta_1 \alpha_1)$ and
$C_{1q_2}(\beta_2 \alpha_2)$ that give the orientation of the
molecular axes \cite{BS}.   Equation (\ref{dd_interaction})
highlights the important point that the orientations of the dipoles
are intimately tied to the relative motion of the dipoles: if a
molecule changes its internal state and sheds angular momentum, that
angular momentum may appear in the orbital motion of the molecules
around each other.

\subsection{Potential matrix elements}

Equation (\ref{dd_interaction}) is a perfectly reasonable way of
writing the classical dipole-dipole interaction.  Quantum
mechanically, however, we are interested in molecules that are in
particular quantum states $|jm {\bar \omega}, \epsilon \rangle$,
rather than molecules whose dipoles point in particular directions
$(\alpha, \beta)$.  We must therefore construct matrix elements of
the interaction potential (\ref{dd_interaction}) in the basis we have
described in Sec. 3.3.

Writing the interaction in the form above has the advantage that
each term in the sum factors into three pieces: one depending on the
coordinates of molecule 1, another depending on the coordinates of
molecule 2, and a third depending on the relative coordinates
$(\theta, \phi)$.  This makes it easier to evaluate the
Hamiltonian in a given basis.  For two molecules we consider the
basis functions
\begin{equation}
\label{two_molecules_basis} \langle \alpha_1 \beta_1 |jm_1 {\bar
\omega}, \epsilon_1 \rangle \langle \alpha_2 \beta_2 |j m_2 {\bar
\omega}, \epsilon_2 \rangle,
\end{equation}
as defined above.  In this basis, matrix elements of the interaction
become
\begin{eqnarray}
\label{dipole_dipole_surface} && \langle jm_1 {\bar \omega},
\epsilon_1 ; jm_2 {\bar \omega}, \epsilon_2 | V_d(\theta, \phi) |
jm^{\prime}_1 {\bar \omega}, \epsilon_1^{\prime} ;
jm^{\prime}_2 {\bar \omega}, \epsilon_2^{\prime} \rangle = \nonumber \\
&& -{ \sqrt{30} \mu^2 \over R^3}   \left(
\begin{array}{ccc} 2 & 1 & 1 \\ q & -q_1 & -q_2 \end{array}
\right)  \\
&& \times \langle jm_1 {\bar \omega}, \epsilon_1 | C_{1q_1} |
jm^{\prime}_1 {\bar \omega}, \epsilon_1^{\prime} \rangle \langle
jm_2 {\bar \omega}, \epsilon_2 | C_{1q_2} | jm^{\prime}_2 {\bar
\omega}, \epsilon_2^{\prime} \rangle C_{2-q}(\theta, \phi) \nonumber
\end{eqnarray}
where matrix elements of the form $\langle jm {\bar \omega},
\epsilon | C_{1q} | jm^{\prime} {\bar \omega}, \epsilon^{\prime}
\rangle$ are evaluated in Eq. (\ref{C_matrix_elements_dressed}).
Conservation of angular momentum projection constrains the values of
the summation indices, so that $q_1 = m_1 - m^{\prime}_1$, $q_2 =
m_2 - m^{\prime}_2$, and $q=q_1 + q_2$ $=(m_1 + m_2) -
(m^{\prime}_1+m^{\prime}_2)$.  To make this model concrete, we
report here the values of the second-rank reduced spherical
harmonics \cite{BS}:
\begin{eqnarray}
C_{20} &=& {1 \over 2} \left( 3\cos^2 \theta -1 \right) \nonumber \\
C_{2 \pm 1} &=& \mp \left( { 3\over 2} \right)^{1/2} \cos \theta
\sin \theta e^{\pm i \phi} \nonumber \\
C_{2 \pm 2} &=& \left( {3 \over 8} \right)^{1/2} \sin^2 \theta
e^{\pm 2i \phi}. \nonumber
\end{eqnarray}
We also tabulate the relevant 3-$j$ symbols in Table III.

Viewed roughly as a collision process, we can think of two molecules
approaching each other with angular momenta $m_1$ and $m_2$,
scattering, and departing with angular momenta $m^{\prime}_1$ and
$m^{\prime}_2$, in which case $q$ is the angular momentum
transferred to the relative angular momentum of the pair of
molecules.  Remarkably, apart from a numerical factor that can be
easily calculated, the part of the quantum mechanical dipole-dipole
interaction corresponding to angular momentum transfer $q$ has an
angular dependence given simply by the multipole term $C_{2-q}$.

Suppose that the molecules, when far apart, are in the well-defined
states (\ref{two_molecules_basis}). Then the diagonal matrix element
of the dipole-dipole potential evaluates to
\begin{equation}
\label{diagonal_interaction}  \left( \mu {m_1 \epsilon_1 {\bar
\omega} \over j(j+1)} \cos 2\delta_{m_1} \right) \left( \mu {m_2
\epsilon_2 {\bar \omega} \over j(j+1)} \cos 2\delta_{m_2} \right)
 { \left( 1 - 3\cos^2 \theta \right) \over R^3}.
\end{equation}
This has exactly the form of the interaction for classical,
polarized dipoles, as in Eq. (\ref{pure_dipole_interaction}).  The
difference is that each dipole $\mu$ is replaced by a
quantum-corrected version (in large parentheses). It is no
coincidence that this is the same quantum-corrected dipole moment
that appeared in the expression (\ref{aligned_dipole}) for the field
due to a single dipole.  When  both dipoles are aligned with the
field, we have $m_1 \epsilon_1 > 0$ and $m_2 \epsilon_2 > 0$ (e.g.,
both molecules are of type $|a \rangle$), and the interaction has
the angular dependence $\propto (1-3\cos^2 \theta)$.  On the other
hand, when one dipole is aligned with the field and the other is
against (e.g., one molecule is of type $|a \rangle$ and the other is
of type $|c \rangle$), then the opposite sign occurs -- just as we
would expect from classical intuition.

\begin{table}
\caption{ The 3-$j$ symbols  needed to construct the matrix
elements in (\ref{dipole_dipole_surface}).  Note that these symbols
remain invariant under interchanging the indices $q_1$ and $q_2$, as well
as under simultaneously changing the signs of $q-1$, $q_2$, and $q$
\cite{BS}. } \vskip 0.2in \center
\begin{tabular}{|c c c | c |}
$q$ & $q_1$ & $q_2$ & $\left( \begin{array}{ccc} 2 & 1 & 1 \\
q &
-q_1 & -q_2 \end{array} \right)$ \\
\hline
0 & 0 & 0 & $\sqrt{ 2/15 }$ \\
0 & 1 & -1 & $1 / \sqrt{30}$ \\
1 & 1 & 0 & $-1 / \sqrt{10}$ \\
2 & 1 & 1 & $1 / \sqrt{5}$
\end{tabular}
\end{table}

More generally, at finite electric field, or at finite values of
$R$,  the molecules do {\it not} remain in the separated-molecule
eigenstates (\ref{two_molecules_basis}), since they exert torques on
one another. The interaction among several different internal
molecular states makes the scattering of two molecules a
``multichannel problem,'' the formulation and solution of which is
described in Chapters XXX.  However, a good way
to visualize the action of the dipole-dipole potential on the
molecules is to construct an adiabatic surface.  To do so, we
diagonalize the interaction at a fixed value of ${\vec R}$, the
relative location of the two molecules.

Before doing this, we must consider the quantum statistics of the
molecules.  If the two molecules under consideration are identical
bosons or identical fermions, then the total two-molecule wave
function must account for this fact. This total wave function is
\begin{eqnarray}
\langle \alpha_1 \beta_1 \gamma_1 |jm_1{\bar \omega} \epsilon_1
\rangle \langle \alpha_2 \beta_2 \gamma_2 \rangle |jm_2{\bar \omega}
\epsilon_2 \rangle  F_{j {\bar \omega}; m_1 \epsilon_1 m_2
\epsilon_2} (R, \theta, \phi). \nonumber
\end{eqnarray}
This wave function is either symmetric or antisymmetric under the
exchange of the two particles, which is accomplished by swapping the
internal states of the molecules, while simultaneously exchanging
their center-or-mass coordinates, i.e., by mapping ${\vec R}$ to
$-{\vec R}$:
\begin{eqnarray}
\label{swap} (\alpha_1 \beta_1) & \leftrightarrow & (\alpha_2
\beta_2) \nonumber
\\ R & \rightarrow & R
\\ \theta & \rightarrow & \pi - \theta \nonumber
\\ \phi & \rightarrow & \pi + \phi. \nonumber
\end{eqnarray}
For the molecule's internal coordinates, a wave function with
definite exchange symmetry is given by
\begin{eqnarray} \label{symmetrized_basis}
&& \langle \alpha_1 \beta_1 \gamma_1 |jm_1{\bar \omega} \epsilon_1
\rangle \langle \alpha_2 \beta_2 \gamma_2 \rangle |jm_2{\bar \omega}
\epsilon_2 \rangle_{s}  \nonumber \\
&& = {1 \over \sqrt{ 2(1 + \delta_{m_1m_2}\delta_{\epsilon_1 \epsilon_2}) }} \\
&& \times \left[ \langle \alpha_1 \beta_1 \gamma_1 | jm_1 \omega
\epsilon_1 \rangle \langle \alpha_2 \beta_2 \gamma_2 | jm_2 \omega
\epsilon_2 \rangle + s \langle \alpha_2 \beta_2 \gamma_2 | jm_1
\omega \epsilon_1 \rangle \langle \alpha_1 \beta_1 \gamma_1 | jm_2
\omega \epsilon_2 \rangle \right]. \nonumber
\end{eqnarray}
 The new index $s = \pm 1$ denotes whether the combination
(\ref{symmetrized_basis}) is even or odd under the interchange.  If $s =
+1$, then $F$ must be symmetric under the transformation ${\vec R}
\rightarrow -{\vec R}$ for bosons, and odd under this transformation for
identical fermions.  If $s=-1$, the reverse must hold.

We now have the tools required to consider the form of the
dipole-dipole interaction beyond the ``pure'' dipolar form
(\ref{diagonal_interaction}).  The details of this analysis will
depend on the Schr\"{o}dinger equation to be solved.  In its
fundamental form, the Schr\"{o}dinger equation reads
\begin{eqnarray}
\label{basic_Schrodinger} \left ( -{\hbar ^2 \over 2m_r } \nabla^2 +
V_d + H_S \right) \Psi = E \Psi. \nonumber
\end{eqnarray}
Here $H_S$ stands for the threshold Hamiltonian that includes
$\Lambda$-doubling and electric field interactions, and is assumed
to be diagonal in the basis (\ref{symmetrized_basis}); and $m_r$ is
the reduced mass of the pair of molecules.   In the usual way, we
expand the total wave function $\psi$ as
\begin{eqnarray}
\Psi( R, \theta, \phi) = {1 \over R} \sum_{i^{\prime}}
F_{i^{\prime}}(R ,\theta, \phi) |i^{\prime} \rangle, \nonumber
\end{eqnarray}
where the index $i$  stands for the collective set of quantum
numbers $\{j {\bar \omega}; m_1 \epsilon_1 m_2 \epsilon_2 s\}$.

Inserting this expansion into the Schr\"{o}dinger equation and
projecting onto the ket $\langle i|$ leads to the following set of
coupled equations:
\begin{eqnarray}
\label{3D_Schrodinger}  && -{\hbar^2 \over 2m_r} \left[ {\partial ^2
 \over
\partial R^2} + {1 \over  R^2 \sin \theta} { \partial \over
\partial \theta} \left( \sin \theta { \partial \over \partial
\theta} \right) + {1 \over R^2 \sin^2 \theta} {\partial^2  \over
\partial \phi^2} \right]F_i \nonumber \\
&& + \sum_{i^{\prime}} \langle i | V_d | i^{\prime} \rangle
F_{i^{\prime}} + \langle i | H_S | i \rangle F_i = EF_i. \nonumber
\end{eqnarray}
If we keep $N$ channels $i$, then this represents a set of $N$
coupled differential equations.  We can, in principle, solve
these subject to physical boundary conditions for any bound or
scattering problem at hand.  For visualization, however, we will
find it convenient to reduce these equations to fewer than three
independent variables.  We carry out this task in the following
subsections.

\subsection{Adiabatic potential energy surfaces in two dimensions}

Applying an electric field in the ${\hat z}$ direction establishes
${\hat z}$ as an axis of cylindrical symmetry for the two-body
interaction.  The angle $\phi$ determines the relative orientation
of the two molecules about this axis, thus the interaction cannot
depend on this angle.  To handle this, we include an additional
factor in our basis set,
\begin{eqnarray}
|m_l \rangle = {1 \over \sqrt{ 2\pi}} \exp(im_l \phi). \nonumber
\end{eqnarray}
We then expand the total wave function as
\begin{eqnarray}
 \Psi^{M_{tot}} (R, \theta, \phi) = {1 \over R}
\sum_{i^{\prime}m_{l}^{\prime}} F_{i^{\prime}m_l^{\prime}}(R,\theta)
|m_l^{\prime} \rangle |i^{\prime} \rangle. \nonumber
\end{eqnarray}
In each term of this expression the quantum numbers must satisfy the conservation
requirement for fixed total angular momentum projection, $M_{tot} = m_1 + m_2 +
m_l$.  In addition, applying exchange symmetry to each term requires
that $F_{i,m_l}(R, \pi - \theta) = s(-1)^{m_l}F_{i,m_l}(R,\theta)$
for bosons, and $s(-1)^{m_l+1}F_{i,m_l}(R,\theta)$ for fermions.

Inserting this expansion into the Schr\"{o}dinger equation yields a
slightly different set of coupled equations:
\begin{eqnarray}
\label{2D_Schrodinger} && -{\hbar^2 \over 2m_r} \left[ {\partial^2
\over \partial R^2} + {1 \over R^2 \sin^2 \theta} {\partial \over
\partial \theta} \left( \sin \theta {\partial \over \partial \theta}
\right) \right]F_{i,m_l}  \\
&& + {\hbar^2 m_l^2 \over 2m_r R^2 \sin^2 \theta} F_{im_l} +
\sum_{i^{\prime}} \langle i | V_d^{2D} | i^{\prime} \rangle
F_{i^{\prime}m_l^{\prime}} + \langle i | H_S | i \rangle F_{im_l} =
EF_{im_l}. \nonumber
\end{eqnarray}
This substitution has the effect of replacing the differential form
of the azimuthal kinetic energy, $\propto \partial^2/\partial
\phi^2$, by an effective centrifugal potential $\propto
m_l^2/R^2\sin^2\theta$.  In addition, the matrix elements of the
dipolar potential $V_d^{2D}$ are slightly different from those of
$V_d$. Recall that the $\theta$ dependent part of the matrix element
(\ref{dipole_dipole_surface}) is proportional to
$C_{2-q}(\theta,\phi)$, which we will write explicitly as
\begin{eqnarray}
C_{2-q}(\theta,\phi) \equiv C_{2-q}(\theta)\exp(-iq\phi). \nonumber
\end{eqnarray}
This equation explicitly defines a new function $C_{2-q}(\theta)$
that is a function of $\theta$ alone, and that is proportional to an
associated Legendre polynomial.  The matrix element of the potential
now includes the following integral:
\begin{eqnarray}
\langle m_l | C_{2-q} | m_l^{\prime} \rangle &=& \int d\phi {1 \over
\sqrt{2 \pi}} e^{-im_l\phi} C_{2-q}(\theta) e^{-iq\phi} {1 \over
\sqrt{2\pi}} e^{im_l^{\prime}\phi} \nonumber \\
&=& {C_{2-q}(\theta) \over 2 \pi} \int d\phi
e^{i(M_{tot}^{\prime}-M_{tot})\phi}
\nonumber \\
&=& \delta_{M_{tot}^{}M_{tot}^{\prime}} C_{2-q}(\theta), \nonumber
\end{eqnarray}
which establishes the conservation of the projection of total
angular momentum by the dipole-dipole interaction.  Therefore,
matrix elements of $V_d^{2D}$ in this representation are identical
to those in of $V_d$ in Eq. (\ref{dipole_dipole_surface}) {\it except}
that the factor $\exp(-iq\phi)$ is replaced by
$\delta_{M_{tot}^{}M_{tot}^{\prime}}$.

With these matrix elements in hand, we can construct solutions to
the coupled differential equations (\ref{2D_Schrodinger}).
However, to understand the character of the potential surface, it is
useful to construct adiabatic potential energy surfaces.  This means
that, for a fixed relative position of the molecules $(R,\theta)$,
we find the energy spectrum of (\ref{2D_Schrodinger}) by
diagonalizing the Hamiltonian $V_c^{2D} + V_d^{2D} + H_S$, where
$V_c^{2D}$ is a shorthand notation for the centrifugal potential discussed
above.  This approximation is common throughout atomic and molecular
physics, and amounts to defining a single surface that comes as
close as possible to representing what is, ultimately, multichannel
dynamics.

\subsection{Example: $j=1/2$ molecules}

Analytic results for the adiabatic surfaces are rather difficult to
obtain.  Consider the simplest realization of our model, a molecule
with spin $j=1/2$. In this case each molecule has four internal
states (two values of $m$ and two values of $\epsilon$), so that the
two-molecule basis comprises sixteen elements.  Dividing these
according to exchange symmetry of the molecules' internal
coordinates, there are ten channels within the manifold of $s=+1$
channels, and six within the $s=-1$ manifold.  These are the cases
we will discuss in the following, although the same qualitative
features also appear in higher-$j$ molecules.

As the simplest illustration of the influence of internal structure
on the dipolar interaction, we will focus on the lowest-energy
adiabatic surface, and show how it differs from the ``pure dipolar''
result (\ref{diagonal_interaction}) as the molecules approach one
another. The physics underlying this difference arises from the fact
that the dipole-dipole interaction becomes stronger as the molecules
get closer together, and at some point this interaction is stronger
than the action of the external field that holds their orientation
fixed in the lab.  The intermolecular distance at which this happens
can be approximately calculated by setting the two interactions equal,
$\mu^2/R_0^3 = \sqrt{(\mu {\cal E})^2 + (\Delta/2)^2}$, yielding a
characteristic distance
\begin{eqnarray}
R_0 = \left( {\mu^2 \over {\sqrt{(\mu {\cal E})^2 + (\Delta/2)^2}}}
\right)^{1/3}. \nonumber
\end{eqnarray}
When $R \gg R_0$, the electric field interaction is dominant, the
dipoles are aligned, and the interaction is given by Eq.
(\ref{diagonal_interaction}). When $R$ becomes comparable to, or
less than, $R_0$, then the dipoles tend to align in a
head-to-tail orientation to minimize their energy, regardless of
their relative location.

Before proceeding, it is instructive to point out how large the
scale $R_0$ can be for realistic molecules.  For the OH molecule
considered above, with $\mu = 1.7$ Debye and $\Delta = 0.06$
cm$^{-1}$, the molecule can be polarized in a field of ${\cal E}
\approx 1600$ V/cm.  At this field, the characteristic radius is
approximately $R_0 \approx 120$ $a_0$ (where $a_0 = 0.053$ nm is the
Bohr radius), far larger than the scale of the molecules themselves.
Therefore, while the dipole-dipole interaction is by far the largest
interaction energy at large $R$, over a significant range of this
potential does not take the usual dipolar form.  To take an even
more extreme case, the molecule NiH has a ground state of $^2\Delta$
symmetry with $j=5/2$ \cite{BC}.  Because it is a $\Delta$, rather
than a $\Pi$, state, its $\Lambda$ doublet is far smaller, probably
on the order of $\sim 10^{-5}$ cm$^{-1}$.  This translates into a
critical field of ${\cal E} \approx 0.5$ V/cm, and a characteristic
radius at this field $R_0 \approx 2000$ $a_0$ $\approx 0.09 $
$\mu$m.  This length is approaching a non-negligible fraction of the
interparticle distance in a Bose-Einstein condensed sample of such
molecules (assuming a density of $10^{14}$ cm$^{-3}$, this spacing
is of order 0.2 $\mu$m). Deviations from the simple dipolar behavior
may thus influence the macroscopic properties of a quantum
degenerate dipolar gas.

As an example, we present in Figure 2 sections of the lowest-energy
adiabatic potential energy surfaces for a ficticious $j=1/2$
molecule whose mass, dipole moment, and $\Lambda$ doublet are equal to those
of OH. These were calculated in a strong-field limit with ${\cal
E}=10^4$ V/cm. Each row corresponds to a particular intermolecular
separation, which is compared to the characteristic radius $R_0$.
However, as noted above, there are two possible ways for the
molecule to have its lowest energy, as illustrated by parts (a) and
(b) of Figure 1. Interestingly, it turns out that these give rise to
rather different adiabatic surfaces.  To illustrate this, we show in
the left column of Figure 2 the surface for a pair of type $|a
\rangle$ molecules, which corresponds at infinitely large $R$ to the
channel $|{1 \over 2} +, {1 \over 2} +; s=1 \rangle$; and in the
right column we show combinations of one type $|a \rangle$ and one
type $|b \rangle$ molecule.  In the latter case, there are two
possible symmetries corresponding to $s=\pm 1$, both of which are
shown. Finally, for comparison, the unperturbed ``pure dipole''
result is shown in all panels as a dotted line.

\begin{figure}
\caption{Angular dependence of adiabatic potential energies for
various combinations of molecules at different interparticle
spacings $R$, which are indicted on the right side.  Dotted lines:
diagonal matrix element of the interaction, assuming both molecules
remain strongly aligned with the electric field.  Solid and dashed
lines: adiabatic surfaces. These surfaces are based on the $j=1/2$
model discussed in the text, using $\mu=1.68$ Debye, $\Delta =
0.056$ cm$^{-1}$, $m_r = 8.5$ amu, and ${\cal E} = 10^4$ V/cm,
yielding $R_0 \sim 70 a_0$. The left hand column presents results for
two molecules of type $|a \rangle$, as labeled in Fig. 1;  in the
right column are results for one molecule of type $|a \rangle$ and
one of type $|b \rangle$, which necessitates specifying an exchange
symmetry $s$.} \centering
\includegraphics[width=0.9\textwidth]{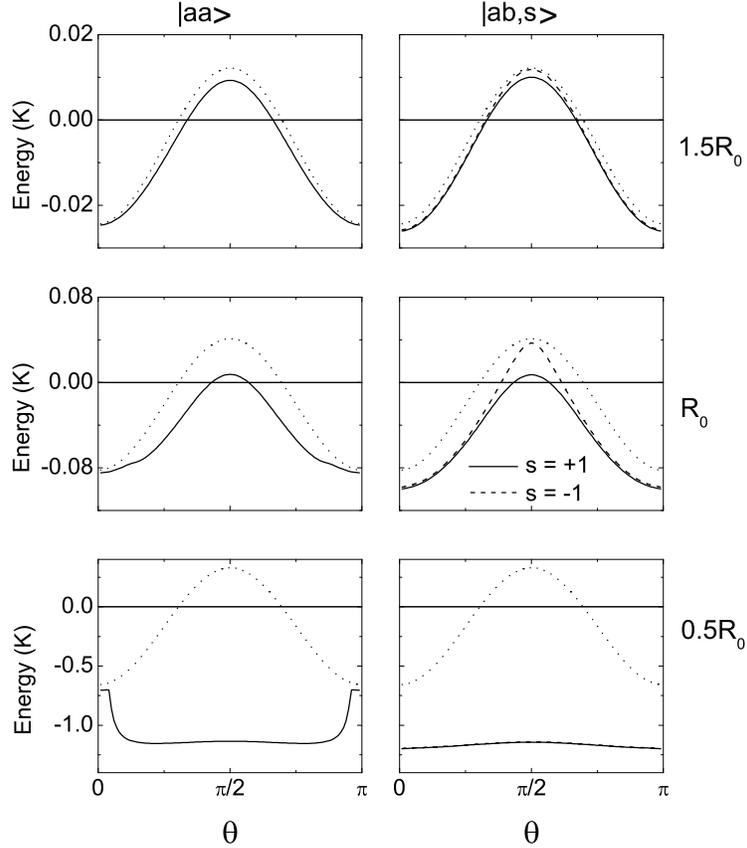}
\end{figure}

Consider first two molecules of type $|a \rangle$ (left column of
Fig.2). For large distances $R>R_0$ (top panel), the adiabatic
potential deviates only slightly from the pure-state result, reducing the
repulsion at $\theta = \pi/2$. When $R$ approaches the
characteristic radius $R_0$ (middle panel), the effect of mixing in
the higher-energy channels becomes apparent. Finally, when $R<R_0$
(lower panel), the mixing is even more significant.  In this case the
dipole-dipole interaction is the dominant energy, with the threshold
energies serving as a small perturbation.  As a consequence, the
quantum numbers $|{1 \over 2} +, {1 \over 2} +; s=1 \rangle$ can no
longer identify the channel.  It is beyond the scope of
this chapter to discuss the corresponding eigenstates in detail.
Nevertheless, we find that for $R<R_0$ the channel $|aa\rangle$
(repulsive at $\theta = \pi/2$) is strongly mixed with the channel
$|ac\rangle$ (attractive at $\theta = \pi/2$).  The combination is
just sufficient that the two channels nearly cancel out one
another's $\theta$-dependence.  At the ends of the range, however,
the adiabatic curve is contaminated by a small amount of channels
containing centrifugal energy $\propto 1/\sin^2\theta$.

The right column of Figure 2 shows adiabatic curves for the mixed
channels, one molecule of type $|a \rangle$ and one of type $|b
\rangle$. In this case there are two possible signs of $s$; These
are distinguished by using solid lines for channel $|{1 \over 2} +,
-{1 \over 2} -; s=1 \rangle$; and dashed lines for channel $|{1
\over 2} +, -{1 \over 2} -; s=-1 \rangle$.  Strikingly, these
surfaces are different both from one another, and from the surfaces
in the left column of the figure.  Ultimately this arises from
different kinds of channel couplings in the potentials
(\ref{dipole_dipole_surface}). Note that, while type $|a \rangle$
and type $|b \rangle$ molecules are identical in their interaction
energy with the electric field, they still represent different
angular momentum states. Nevertheless, molecules in these channels
still closely reflect the pure dipolar potential at large $R$, and
become nearly $\theta$-independent for small $R$.

A further important point is that the potentials described here
represent large energies as compared to the mK or $\mu$K
translational kinetic energies of cold molecules, and will therefore
significantly influence their dynamics. Further, the potentials
depend strongly on the value of the electric field of the
environment, both through the direct effect of polarization on the
magnitude of the dipole moments, and through the influence of the
field on the characteristic radius $R_0$.  It is this sensitivity to
field that opens the possibility of control over interactions in an
ultracold dipolar gas.

Although we have limited the discussion here to the lowest adiabatic
state, interesting phenomena are also expected to arise due to
avoided crossings in excited states. Notable is a collection of
long-range quasi-bound states, whose intermolecular spacing is
roughly centered around $R_0$ \cite{Avdeenkov}.  Such states could
conceivably be used to associate pairs of molecules into
well-characterized transient states, furthering the possibilities of
control of molecular interactions.

\subsection{Adiabatic potential energy curves in one dimension: partial waves}

For many scattering applications, it is not necessarily convenient
to express the dipole-dipole interaction as a surface (more
properly, a set of surfaces) in the variables $(R,\theta,\phi)$
describing the relative position and orientation of the molecules.
Rather, it is useful to expand the relative angular coordinates
$(\theta,\phi)$ in a basis as well.
To do this, the basis set (\ref{two_molecules_basis}) is
augmented by spherical harmonics describing the relative orientation
of the molecules, to become
\begin{eqnarray}
\langle \alpha_1 \beta_1 \gamma_1|jm_1{\bar \omega} \epsilon_1
\rangle \langle \alpha_2 \beta_2 \gamma_2 |jm_2 {\bar \omega}
\epsilon_2 \rangle \langle \theta \phi |lm_l \rangle, \nonumber
\end{eqnarray}
with
\begin{eqnarray}
\langle \theta \phi |lm_l \rangle = Y_{lm_l}(\theta \phi) = \sqrt{
{2l+1 \over 4 \pi} } C_{lm_l}(\theta \phi). \nonumber
\end{eqnarray}
The total wave function is therefore described by the superposition
\begin{eqnarray}
\Psi^{M_{tot}}(R,\theta,\phi) = {1 \over R}
\sum_{i^{\prime},l^{\prime},m_l^{\prime}}
F_{i^{\prime},l^{\prime},m_l^{\prime}}
Y_{l^{\prime}m_l^{\prime}}(\theta\phi) |i^{\prime} \rangle,
\nonumber
\end{eqnarray}
which represent a conventional expansion into partial waves.
The wave function is, as above, restricted by conservation of
angular momentum to require $M_{tot} = m_1 + m_1 + m_l$ to have a
constant value. Moreover, the effect of the symmetry operation
(\ref{swap}) on $|lm_l\rangle$ is to introduce a phase factor
$(-1)^l$. Therefore, the wave function is restricted to $s(-1)^l =
1$ for bosons, and $s(-1)^l = -1$ for fermions.

The effect of this extra basis function is to replace the $C_{2-q}$
factor in (\ref{dipole_dipole_surface}) by its matrix element
\begin{eqnarray}
\label{C_matrix_element} C_{2-q} \rightarrow && \langle lm_l |
C_{2-q} | l^{\prime}m_l^{\prime}
\rangle  \\
&& = \sqrt{ (2l+1)(2l^{\prime}+1) } (-1)^{m_l} \left(
\begin{array}{ccc} l & 2 & l \\ 0 & 0 & 0 \end{array} \right) \left(
\begin{array}{ccc} l & 2 & l \\ -m_l & -q & m_l^{\prime} \end{array}
\right) \nonumber
\end{eqnarray}
From this expression it is seen that the angular momentum $q$, lost
to the internal degrees of freedom of the molecules, appears as the
change in their relative orbital angular momentum.  From here, the
effects of field dressing are exactly as treated above.

The quantum number $l$, as is usual in quantum mechanics when
treated in spherical coordinates, represents the orbital angular
momentum of the pair of molecules about their center of mass.
Following the usual treatment, this leads to a set of coupled radial
Schr\"{o}dinger equations for the relative motion of the molecules:
\begin{eqnarray}
-{\hbar^2 \over 2m_r} {d^2F_{ilm_l} \over dR^2} &&+ {\hbar^2 l(l+1)
\over 2m_r R^2} F_{ilm_l} \nonumber \\
&& + \sum_{i^{\prime}} \langle i | V_d^{1D} | i^{\prime} \rangle
F_{i^{\prime}l^{\prime}m_l^{\prime}} + \langle i | H_S | i \rangle
F_{ilm_l} = EF_{ilm_l}. \nonumber
\end{eqnarray}
The second term in this expression represents a centrifugal
potential $\propto 1/R^2$, which is present for all partial waves
$l>0$.

\subsection{Asymptotic form of the interaction}

Casting the Schr\"{o}dinger equation as an expansion in partial
waves, and the interaction as a set of curves in $R$, rather than as
a surface in $(R,\theta, \phi)$, allows us to explore more readily
the long-range behavior of the dipole-dipole interaction.

The first 3-$j$ symbol in (\ref{C_matrix_element}) vanishes unless
$l + 2 + l^{\prime}$ is an odd number, meaning that even(odd)
partial waves are coupled only to even(odd) partial waves. Moreover,
the values of $l$ and $l^{\prime}$ can differ by at most two.  Thus
the dipolar interaction can change the orbital angular momentum
state from $l=2$ to $l^{\prime}=4$, for instance, but not to
$l^{\prime}=6$. Finally, the interaction vanishes altogether for
$l=l^{\prime}=0$, meaning that the dipole-dipole interaction
nominally vanishes in the $s$-wave channel.

Since all other channels have higher energy, due to their
centrifugal potentials, it appears that the lowest adiabatic curve
is trivially equal to zero.  This is not the case, however, since
the $s$-wave channel is coupled to a nearby $d$-wave channel with
$l=2$.  Ignoring for the moment higher partial waves, the
Hamiltonian corresponding to a particular channel $i$ at long range
has the form
\begin{eqnarray}
\left( \begin{array}{cc} 0 & A_{02} / R^3 \\ A_{20} / R^3 & 3
\hbar^2 / m_r R^2 + A_{22} / R^3
\end{array}\right). \nonumber
\end{eqnarray}
Here $A_{02}=A_{20}$ and $A_{22}$ are coupling coefficients that
follow from the expressions derived above.  Note that all these
coefficients are functions of the electric field ${\cal E}$.  Now,
the comparison of dipolar and centrifugal energies defines another
typical length scale for the interaction, namely, the one where
$\mu^2 / R^3 = \hbar^2 / m_r R^2$, defining a ``dipole radius'' $R_D
= \mu^2 m_r / \hbar^2$.  (More properly, one could define an
electric-field-dependent radius by substituting $A_{02}$ for
$\mu^2$).  For OH, this length is $\sim 6800$ $a_0$, while for NiH
it is $\sim 9000$ $a_0$.

When the molecules are far apart, $R>R_D$, the dipolar interaction
is a perturbation.  The size of this perturbation on the $s$-wave
interaction is found through second-order perturbation theory to be
\begin{eqnarray}
{ (A_{02}/R^3)^2 \over 3\hbar^2 / m_r R^2} \sim \left({A_{02}^2 m_r
\over 3\hbar^2} \right) {1 \over R^4}. \nonumber
\end{eqnarray}
Therefore, at very large intermolecular distances, the
effective potential, as described by the lowest adiabatic curve,
carries a $1/R^4$ dependence on $R$, rather than the nominal $1/R^3$
dependence.  At closer range, when $R<R_D$, the $1/R^3$ terms
dominate the $1/R^2$ centrifugal interaction, and the potential
reduces again to the expected $1/R^3$ dependence on $R$.

\section{Acknowledgements} I would like to acknowledge many
fruitful discussions of molecular dipoles over the years, notably
those with Aleksandr Avdeenkov, Doerte Blume, Daniele Bortolotti,
Jeremy Hutson, and Chris Ticknor.  This work was supported by the
JILA NSF Physics Frontier Center.


\begin{thebibliography}{20}
\bibitem{BS} D. M. Brink and G. R. Satchler, {\it Angular Momentum}
(Oxford University Press, 3rd Edition, 1993).
\bibitem{BC} J. Brown and A. Carrington, {\it Rotational
Spectroscopy of Diatomic Molecules} (Cambridge University Press,
2003).
\bibitem{Jackson} J. D. Jackson, {\it Classical Electrodynamics}
(Wiley, New York, 2nd Edition, 1975).
\bibitem{Bethe_Salpeter} H. A. Bethe and E. E. Salpeter, {\it
Quantum Mechanics of One- and Two-Electron Atoms} (Plenum Press,
1977).
\bibitem{Hougen} J. T. Hougen, ``The Calculation of Rotational
Energy  Levels and Rotational Line Intensities in Diatomic
Molecules,'' NBS Monograph 115, availalbe online at
http://physics.nist.gov/Pubs/Mono115/
\bibitem{Bloch} L. Allen and J. H. Eberly, {\it Optical Resonance
and Two-Level Atoms} (Wiley, New York, 1975).
\bibitem{Avdeenkov} A. V. Avdeenkov, D. C. E. Bortolotti, and J. L.
Bohn, ``Field-Linked States of Ultracold Polar Molecules,'' Phys.
Rev. A {\bf 69}, 012710 (2004).
\end{thebibliography}
\end{document}